\documentclass[12pt]{article}

\usepackage{geometry,bm}
\usepackage{amsmath}
\usepackage{amssymb}
\usepackage[english]{babel}
\usepackage[latin1]{inputenc}
\usepackage{mathrsfs, amssymb}
\usepackage{array}
\usepackage{graphicx}
\usepackage{xcolor}
\usepackage{subfigure}
\usepackage[round]{natbib}
\usepackage[bottom]{footmisc}
\usepackage{theorem}
\theoremstyle{break}

\newtheorem{definition}{Definition}

\numberwithin{equation}{section} 

\newcommand{\bs}[1]{\boldsymbol{#1}}
\newcommand{\mc}[1]{\mathcal{#1}}

\begin{document}
\title{Instrumental Variable Bayesian Model Averaging via Conditional Bayes Factors}

\author{Anna Karl and Alex Lenkoski\footnote{\noindent \textit{Corresponding author address:} Alex Lenkoski, Institute of Applied Mathematics, Heidelberg University, Im Neuenheimer Feld 294, 69120 Heidelberg, Germany
\newline{E-mail: alex.lenkoski@uni-heidelberg.de}}\\
\textit{Heidelberg University, Germany}}
\maketitle
\linespread{1}
\begin{abstract}
We develop a method to perform model averaging in two-stage linear regression systems subject to endogeneity. Our method extends an existing Gibbs sampler for instrumental variables to incorporate a component of model uncertainty. Direct evaluation of model probabilities is intractable in this setting. We show that by nesting model moves inside the Gibbs sampler, model comparison can be performed via conditional Bayes factors, leading to straightforward calculations.  This new Gibbs sampler is only slightly more involved than the original algorithm and exhibits no evidence of mixing difficulties. We conclude with a study of two different modeling challenges: incorporating uncertainty into the determinants of macroeconomic growth, and estimating a demand function by instrumenting wholesale on retail prices.
\vspace{.05cm}
\\

\end{abstract}
\linespread{1.3}
\pagebreak
\section{Introduction}
\thispagestyle{empty}
\indent We consider the problem of incorporating instrument and covariate uncertainty into the Bayesian estimation of an instrumental variable (IV) regression system.  The concepts of model uncertainty and model averaging have received widespread attention in the economics literature for the standard linear regression framework (see, e.g. \citet{fernandez_et_2001}, \citet{eicher_et_2007} and references therein).  However, these frameworks do not directly address endogeneity and only recently has attention been paid to this important component.  Unfortunately, the nested nature of IV estimation renders direct model comparison exceedingly difficult.\\
\indent This has led to a number of different approaches. \citet{durlauf_et_2008}, \citet{cohen-cole_et_2009} and \citet{durlauf_et_2011} consider approximations of marginal likelihoods in a framework similar to two-stage least squares. \citet{lenkoski_et_2011} continues this development with the two-stage Bayesian model averaging (2SBMA) methodology, which uses a framework developed by \citet{kleibergen_zivot_2003} to propose a two-stage extension of the unit information prior \citep{kass_wasserman_1995}.  Similar approaches in closely related models have been developed by \citet{morales-benito_2009} and \citet{chen_et_2009}.\\
\indent \citet{koop_et_2012} develop a fully Bayesian methodology that does not utilize approximations to integrated likelihoods. They develop a reversible jump Markov chain Monte Carlo (RJMCMC) algorithm \citep{green_1995}, which extends the methodology of \citet{holmes_et_2002}.  The authors then show that the method is able to handle a variety of priors, including those of \citet{dreze_1976}, \citet{kleibergen_vandijk_1998} and \citet{strachan_inder_2004}.  However, the authors note that direct application of RJMCMC leads to significant mixing difficulties and rely on a complicated model move procedure that has similarities to simulated tempering to escape local model modes.\\
\indent We propose an alternative solution to this problem, which we term Instrumental Variable Bayesian Model Averaging (IVBMA).  Our method builds on a Gibbs sampler for the IV framework, discussed in \citet{rossi_et_2006}.  While direct model comparisons are intractable, we introduce the notion of a conditional Bayes factor (CBF), first discussed by \citet{dickey_gunel_1978}.  The CBF compares two models in a nested hierarchical system, conditional on parameters not influenced by the models under consideration.  We show that the CBF for both first and second-stage models is exceedingly straightforward to calculate and essentially reduces to the normalizing constants of a multivariate normal distribution.\\
\indent This leads to a procedure in which model moves are embedded in a Gibbs sampler, which we term MC3-within-Gibbs.  Based on this order of operations, IVBMA is then shown to be only trivially more difficult than the original Gibbs sampler that does not incorporate model uncertainty.  A three-step procedure is updated to a five-step procedure and as such, IVBMA appears to have limited issues regarding mixing.  Furthermore, the routines discussed here are contained in the {\tt R} package {\tt ivbma}, which can be freely downloaded from the Comprehensive {\tt R} Archive Network ({\tt CRAN}).\\
\indent The article proceeds as follows.  The basic framework we consider, and the Gibbs sampler ignoring model uncertainty is discussed in Section 2. Section 3 reviews the concept of model uncertainty, introduces the notion of CBFs and derives the conditional model probabilities used by IVBMA.  In Section 4 we conclude with two data analyses.  The first is the classic problem of modeling uncertainty in macroeconomic growth determinants, which has proven a testing-ground for BMA in economics.  Second, we consider the problem of modeling an uncertain demand function, in particular the volume of demand for margarine in Denver, Colorado, between January 1993 and March 1995. In Section 5 we conclude.  Appendices give details of the calculations outlined in Sections 2 and 3.  

\section{Methodology}
\thispagestyle{empty}
\subsection{Description of the Model}
We consider the classic, two-stage endogenous variable model:
\begin{align}
\textbf{Y}&=\bs{X}\beta+\textbf{\textit{W}}\bs{\gamma}+\bs{\epsilon} \label{equ1}\\
\bs{X}&=\textbf{\textit{Z}}\boldsymbol{\delta}+\textbf{\textit{W}}\boldsymbol{\tau}+\bs{\eta} \label{equ2}
\end{align}
with 
\begin{equation}\label{err} \begin{pmatrix}\epsilon_i \\ \eta_i \end{pmatrix}\thicksim \mathcal{N}_{2}(0,\boldsymbol{\Sigma}) \end{equation} and  \begin{equation*} \boldsymbol{\Sigma}=\begin{pmatrix}\sigma_{11} &\sigma_{12} \\ \sigma_{21} &\sigma_{22}\end{pmatrix};\;\sigma_{12}=\sigma_{21}\neq 0. 
\end{equation*}\medskip
\indent In what follows we restrict the response variable $\bs{Y}$ and the endogenous explanatory factor $\bs{X}$ to be $n\times 1$ \footnote{In the Conclusions section we outline the straightforward steps necessary to incorporate multiple endogenous variables.}. $\textbf{\textit{W}}$ denotes an $n\times p$ matrix of further explanatory variables with $p \times 1$ parameter vectors $\bs{\gamma}$ and $\bs{\tau}$. The instrumental variables are described by the $n\times q$ matrix $\bs{Z}$ with $\bs{\delta}$ a $q\times 1$ parameter vector. The coefficient $\beta$ is a scalar.\\
\indent Due to \eqref{err} the error terms are homoscedastic and correlated in each component, since $\text{Cov}(\epsilon,\eta)= \sigma_{12} = \sigma_{21} \neq 0$ for all observations, requiring joint estimation of the system (\ref{equ1}) and (\ref{equ2}) in order to draw appropriate inference for the parameters in the outcome equation (\ref{equ1}).  The assumption of bivariate normality in $(\ref{err})$ is helpful in deriving a fast algorithm for posterior determination; in the Conclusions section we discuss how this may be relaxed.
\subsection{Calculation of the Conditional Posterior Distributions}\label{sec:gibbs}
In this paper we focus solely on the Bayesian estimation of the IV framework detailed above.  We consider a prior framework detailed in \citet{rossi_et_2006}--extended to the multivariate setting--as it lends itself to quick posterior estimation through Gibbs sampling.\\
\indent In order to adequately explain the CBF calculations we perform in Section~\ref{sec:ModSel}, it is helpful to review the derivation of the conditional posterior distributions.  The following three subsections will present the calculation of the posterior distribution $pr(\bs{\theta}|\mathcal{D})$ of the parameter vector $\bs{\theta}=(\beta,\boldsymbol{\gamma},\boldsymbol{\delta},\boldsymbol{\tau},\boldsymbol{\Sigma})$ of our model \eqref{equ1}-\eqref{err}, conditional on the data $\mathcal{D}=(\bs{Y},\bs{X},\textbf{\textit{W}},\textbf{\textit{Z}})$.\\
\indent The Gibbs sampler we outline below divides $\bs{\theta}$ into three subvectors, $\boldsymbol{\rho}= [\beta\mbox{ }\bs{\gamma}']'$,  $\boldsymbol{\lambda}=[\boldsymbol{\delta}'\mbox{ }\boldsymbol{\tau}']'$ and $\boldsymbol{\Sigma}$
with $\boldsymbol{\rho} \in \mathbb{R}^{1+p}$, $\boldsymbol{\lambda} \in \mathbb{R}^{q+p}$, and $\bs{\Sigma} \in \mathbb{P}_2$, where $\mathbb{P}_2$ denotes the cone of $2 \times 2$ positive definite matrices.  Appendix A gives full details of the conditional distributions derived below.

\subsubsection{Step 1: The Conditional Posterior Distribution of $\bs{\rho}$}\label{sec:gibbs1}
Assuming a standard normal prior for the second stage regressors $\bs{\rho}\thicksim\mathcal{N}(0,\mathbb{I}_{1+p}),$ we have
\begin{equation}\label{distrho} 
\boldsymbol{\rho}|\boldsymbol{\lambda},\boldsymbol{\Sigma},\mathcal{D} \thicksim\mathcal{N}(\boldsymbol{\hat{\rho}},\boldsymbol{\Xi}^{-1}), 
\end{equation}
where $\boldsymbol{\hat{\rho}} = \xi^{-1}\tilde{\bs{Y}}'\textbf{\textit{V}} \bs{\Xi}^{-1}$, $\boldsymbol{\Xi} = \mathbb{I}_{1+p} + \xi^{-1}\bs{V}'\bs{V}$ with $\tilde{\bs{Y}} = \bs{Y} - \frac{\sigma_{21}}{\sigma_{22}}\bs{\eta}$, $\textbf{\textit{V}} = [\bs{X}\hspace{2mm}\textbf{\textit{W}}]$ and $\xi = \sigma_{11} - \frac{\sigma_{21}^2}{\sigma_{22}}$.  Details are given in Appendix A.
\subsubsection{Step 2: The Conditional Posterior Distribution of $\bs{\lambda}$}\label{sec:gibbs2}
Assuming a standard normal prior for the first stage regressors $\boldsymbol{\lambda}\thicksim\mathcal{N}(0,\mathbb{I}_{q+p})$,
we have 
\begin{equation}\label{distlambda} 
\boldsymbol{\lambda}|\boldsymbol{\rho},\boldsymbol{\Sigma},\mathcal{D} \thicksim\mathcal{N}(\hat{\boldsymbol{\lambda}},\boldsymbol{\Omega}^{-1}), 
\end{equation} 
where $\hat{\boldsymbol{\lambda}} = \textbf{S}'\textbf{T}\boldsymbol{\Omega}^{-1},\boldsymbol{\Omega} = \mathbb{I}_{q+p} + \textbf{T}'\textbf{T}$. Here, $\boldsymbol{S}$ is a $2n \times 1$ matrix formed from $\bs{Y},\bs{X}$ and $\bs{\Sigma}$ and $\bs{T}$ is a $2n \times (p + q)$ matrix formed from $\bs{W}$, $\bs{Z}$ and $\bs{\Sigma}$, whose construction is outined in Appendix A.
\subsubsection{Step 3: The Conditional Posterior Distribution of \boldmath$\Sigma$} 
Finally, to determine $pr(\boldsymbol{\Sigma}|\boldsymbol{\rho},\boldsymbol{\lambda},\mathcal{D}),$ we use an inverse-Wishart prior \citep[e.g.][]{anderson_1984}. Thus, $\boldsymbol{\Sigma}\thicksim\mathcal{W}^{-1}(\mathbb{I}_2,3)$ and as the inverse-Wishart is conjugate we have
\begin{equation}\label{distSigma} 
\boldsymbol{\Sigma}|\boldsymbol{\rho},\boldsymbol{\lambda},\mathcal{D} \thicksim\mathcal{W}^{-1}(\mathbb{I}_2+\boldsymbol{Q},3+n) 
\end{equation}
where $\boldsymbol{Q}= [\bs{\epsilon}\mbox{ }\bs{\eta}]'[\bs{\epsilon} \mbox{ } \bs{\eta}]$.
\section{Incorporating Model Uncertainty}\label{sec:ModSel}
We outline our method for incorporating model uncertainty into the estimation of the framework (\ref{equ1}) and (\ref{equ2}).  In order to explain the motivation behind our CBF approach, we first review some basic results from classic model selection problems.  We then show how the concept of Bayes Factors can be usefully embedded in a Gibbs sampler yielding CBFs.  These CBFs are then shown to yield straightforward calculations.  The section concludes with an overview of the full IVBMA procedure.
\subsection{Bayes Factors}
In a general framework, incorporating model uncertainty involves considering a collection of candidate models $\mc{I}$, using the data $\mathcal{D}$. Each model $I$ consists of a collection of probability distributions for the data $\mathcal{D}$, $\{pr(\mathcal{D}|\psi),\psi \in \Psi_{I}\}$ where $\Psi_{I}$ denotes the parameter space for the parameters of model $I$ and is a subset of the full parameter space $\Psi$.\\
\indent By letting the model become an additional parameter to be assessed in the posterior, we aim to calculate the posterior model probabilities given the data $\mathcal{D}$. By Bayes' rule 
\begin{equation}\label{pmp}
 pr(I|\mathcal{D})=\frac{pr(\mathcal{D}|I)pr(I)}{\sum_{I'\in\mc{I}}{pr(\mathcal{D}|I')pr(I')}},
 \end{equation}
where $pr(I)$, denotes the prior probability for model $I\in\mc{I}$.\\
\indent The integrated likelihood $pr(\mathcal{D}|I)$, is defined by
\begin{equation}\label{marglik}
 pr(\mathcal{D}|I)=\int_{\Psi_I} pr(\mathcal{D}|\psi)pr(\psi|I) d\psi,
 \end{equation}
where $pr(\psi|I)$ is the prior for $\psi$ under model $I$, which by definition has all its mass on $\Psi_I$.\\ 
\indent One possibility for pairwise comparison of models is offered by the Bayes factor (BF), which is in most cases defined together with the posterior odds \citep{kass_raftery_1995}.
\begin{definition}[Posterior odds and Bayes factor]
The posterior odds of model $I$ versus model $I'$ is given by
\begin{equation*} 
\frac{pr(I|\mathcal{D})}{pr(I'|\mathcal{D})}=\frac{pr(\mathcal{D}|I)}{pr(\mathcal{D}|I')}\frac{pr(I)}{pr(I')}, 
\end{equation*} where 
\begin{equation*} 
\frac{pr(\mathcal{D}|I)}{pr(\mathcal{D}|I')} \quad\text{and}\quad \frac{pr(I)}{pr(I')}
\end{equation*} 
denote the Bayes factor and the prior odds of $I$ versus $I'$, respectively.
\end{definition}
\indent When the integrated likelihood (\ref{marglik}) and thus the BF can be computed directly, a straightforward method for exploring the model space, Markov Chain Monte Carlo Model Composition (MC3), was developed by \citet{madigan_york_1995}.\\
\indent MC3 determines posterior model probabilities by generating a stochastic process that moves through the model space $\mathcal{I}$ and has equilibrium distribution $pr(I|\mathcal{D})$. Given the current state $I^{(s)}$, MC3 proposes a new model $I'$ according to a proposal distribution $q(\cdot|\cdot)$, calculates
$$
\alpha = \frac{pr(\mathcal{D}|I')pr(I')q(I^{(s)}|I')}{pr(\mathcal{D}|I^{(s)})pr(I^{(s)})q(I'|I^{(s)})}
$$
and sets $I^{(s + 1)} = I'$ with probability $\min\{\alpha,1\}$ otherwise setting $I^{(s + 1)} = I^{(s)}$.\\
\subsection{Model Determination for Two-Staged Problems}\label{sec:cbfs}
We now consider the incorporation of model uncertainty into the system (\ref{equ1}) and (\ref{equ2}).  We follow the notation of \citet{lenkoski_et_2011}.  Associated with the outcome equation $(\ref{equ1})$ we consider a collection of models $\mc{L}$.  Each $L \in \mc{L}$ consists of a different restriction on the parameter $\bs{\rho}$ and we denote $\Gamma_{L} \subset \mathbb{R}^{1 + p}$ this restricted space.   Similarly in the instrument equation (\ref{equ2}) we consider a collection $\mc{M}$ which impose restrictions on the vector $\bs{\lambda}$ and associate with each $M\in \mc{M}$ a space $\Lambda_{M}\subset \mathbb{R}^{p + q}$.\\
\indent Ideally, we would be able to incorporate model uncertainty into this system in a manner analogous to that described above. Unfortunately,
$$
pr(\mc{D}|L, M) = \int_{\mathbb{P}_2}\int_{\Lambda_M}\int_{\Gamma_L}pr(\mc{D}|\bs{\rho},\bs{\lambda},\bs{\Sigma})pr(\bs{\rho}|L)pr(\bs{\lambda}|M)pr(\bs{\Sigma})d\bs{\rho}d\bs{\lambda}d\bs{\Sigma}
$$
cannot be directly calculated in any obvious manner.  Therefore an implementation of MC3 on the product space of $\mc{L} \times \mc{M}$ is infeasible.  What we show below, however, is that embedding MC3 within the Gibbs sampler, and therefore calculation using CBFs to move between models, offers an extremely efficient solution. CBFs were originally discussed in \citet{dickey_gunel_1978} and for the IV framework are defined below.
\begin{definition}[Conditional Bayes factor]                
Given the system (\ref{equ1}) and (\ref{equ2}), let $\boldsymbol{\Sigma}$ be the covariance matrix and $\boldsymbol{\lambda}$ and $\boldsymbol{\rho}$ denote the parameters of the first and second stage, respectively. \\
(a) The CBF of second stage models $L$ and $L'$ is defined as
\begin{equation*} 
  CBF_{sec}=\frac{pr(\mc{D}|L',\boldsymbol{\lambda}, \boldsymbol{\Sigma})}{pr(\mc{D}|L,\boldsymbol{\lambda}, \boldsymbol{\Sigma})}. 
\end{equation*}
(b) For first stage models $M$ and $M'$ the CBF is given by
\begin{equation*} 
  CBF_{fst}=\frac{pr(\mc{D}|M',\boldsymbol{\rho}, \boldsymbol{\Sigma})}{pr(\mc{D}|M,\boldsymbol{\rho}, \boldsymbol{\Sigma})}. 
\end{equation*}
\end{definition} 
Considering $CBF_{sec}$, we can see that it relies on determining the quantity
$$
pr(\mc{D}|L,\boldsymbol{\lambda}, \boldsymbol{\Sigma}) = \int_{\Gamma_L} pr(\mc{D}|\bs{\rho}, \bs{\lambda}, \bs{\Sigma})pr(\bs{\rho}|L)d\bs{\rho}
$$
which is, in essence, an integrated likelihood for model $L$ conditional on fixed values of $\bs{\lambda}$ and $\bs{\Sigma}$.  In Appendix B we show that
\begin{equation}\label{eq:secintlik}
\int_{\Gamma_{L}} pr(\mc{D}|\bs{\rho}, \bs{\lambda}, \bs{\Sigma})pr(\bs{\rho}|L)d\bs{\rho} \propto
|\boldsymbol{\Xi}_{L}|^{-1/2}\exp\left(\frac{1}{2}\hat{\boldsymbol{\rho}}_{L}'\boldsymbol{\Xi}_{L}\hat{\boldsymbol{\rho}}_{L}\right).
\end{equation}
Where $\hat{\bs{\rho}}_{L}$ and $\bs{\Xi}_{L}$ are defined in Appendix B, but are exactly analogous to the $\hat{\bs{\rho}}$ and $\bs{\Xi}$ discussed in section $\ref{sec:gibbs1}$, restricted to the subset of $\bs{X}$ and $\bs{W}$ included in model $L$.\\
\indent Similarly, in Appendix B we show that
\begin{equation}\label{eq:fstintlik}
pr(\mc{D}| M,\boldsymbol{\rho},\boldsymbol{\Sigma}) \varpropto |\boldsymbol{\Omega}_{M}|^{-1/2}\exp\left(\frac{1}{2}\hat{\boldsymbol{\lambda}}_{M}'\boldsymbol{\Omega}_{M}\hat{\boldsymbol{\lambda}}_{M}\right),
\end{equation}
where $\hat{\bs{\lambda}}_{M}$ and $\bs{\Omega}_{M}$ are again defined in Appendix B, but are analogous\footnote{However, as noted in the Appendix, when $\rho_1 = 0$, and thus the endogenous variable is not included in (\ref{equ1}), the update is altered and resembles a seemingly unrelated regression update.} to the similar quantities discussed in Section~\ref{sec:gibbs2}.\\
\indent Equations (\ref{eq:secintlik}) and (\ref{eq:fstintlik}) show that both $CBF_{fst}$ and $CBF_{sec}$ can be calculated directly.  Furthermore, these calculations are extremely straightforward, and involve computing little more than the parameters necessary for sampling in the Gibbs sampler.
\subsection{Model Space Priors}
\indent Setting a prior on models in the IVBMA framework necessitates--at a minimum--some subtlety in order to guarantee the pair constitute an IV specification.  Let $\mc{A} \subset \mc{L}\times\mc{M}$ such that $(L,M)\in \mc{A}$ if and only if $M\setminus L \neq \emptyset$.  We therefore are only interested in considering model pairs in the collection $\mc{A}$.\\
\indent In what follows, we assume
$$
pr(L,M)\propto \bs{1}\{(L,M)\in \mc{A}\}.
$$
In other words, we assume a uniform prior on the space of models in $\mc{A}$.  Other priors on the model space \citep[e.g.][]{brock_et_2003, scott_berger_2006,durlauf_et_2008,ley_steel_2009} could easily be accommodated.
\subsection{The IVBMA Algorithm}
Building upon the original Gibbs sampler discussed in Section~\ref{sec:gibbs}, and the derivations in Section~\ref{sec:cbfs} we now outline the IVBMA algorithm, which relies on an MC3-within-Gibbs sampler\footnote{In reality, this is simply a special case of the Metropolis-within-Gibbs algorithm \citep[see][]{chib_greenberg_1995}, since the MC3 step can be considered a Metropolis-Hastings step in the space of models.}.  IVBMA creates a sequence $\{\bs{\theta}^{(1)}, \dots, \bs{\theta}^{(S)}\}$ where 
$$
\bs{\theta}^{(s)} = \{\bs{\rho}^{(s)}, L^{(s)}, \bs{\lambda}^{(s)}, M^{(s)}, \bs{\Sigma}^{(s)}\}
$$
with $\bs{\rho}^{(s)} \in \Gamma_{L^{(s)}}$, $\bs{\lambda}^{(s)} \in \Lambda_{M^{(s)}}$ and $(L^{(s)}, M^{(s)})\in\mc{A}$.  Given the current state $\bs{\theta}^{(s)}$ and the data $\mc{D}$, IVBMA proceeds as follows 
\begin{itemize}
\item[\textbf{1.}] \textbf{Update $L$:}  First, sample $L'$ from the neighborhood of $L^{(s)}$ (i.e. uniformly on those models that differ from $L^{(s)}$ by only one variable).  Then calculate
$$
\alpha = \frac{pr(\mc{D}|L',\bs{\lambda}^{(s)}, \bs{\Sigma}^{(s)})}{pr(\mc{D}|L^{(s)},\bs{\lambda}^{(s)}, \bs{\Sigma}^{(s)})}\bs{1}\{(L',M^{(s)})\in\mc{A}\}
$$
using Equation (\ref{eq:secintlik}).  With probability $\min\{\alpha,1\}$ set $L^{(s + 1)} = L'$, otherwise set $L^{(s + 1)} = L^{(s)}$.
\item[\textbf{2.}] \textbf{Update $\bs{\rho}$:}  Sample $\bs{\rho}^{(s + 1)} \sim \mathcal{N}(\hat{\bs{\rho}}_{L^{(s + 1)}}, \bs{\Xi}_{L^{(s + 1)}}^{-1})$ as discussed in Appendix B.
\item[\textbf{3.}] \textbf{Update $M$:}  Sample $M'$ from the neigborhood of $M^{(s)}$, then calculate
$$
\alpha = \frac{pr(\mc{D}|M',\bs{\rho}^{(s + 1)}, \bs{\Sigma}^{(s)})}{pr(\mc{D}|M^{(s)},\bs{\rho}^{(s + 1)}, \bs{\Sigma}^{(s)})}\bs{1}\{(L^{(s + 1)}, M')\in\mc{A}\}
$$
using Equation (\ref{eq:fstintlik}).  With probability $\min\{\alpha,1\}$ set $M^{(s + 1)} = M'$, otherwise set $M^{(s + 1)} = M^{(s)}$.
\item[\textbf{4.}] \textbf{Update $\bs{\lambda}$:} Sample $\bs{\lambda}^{(s + 1)} \sim \mc{N}(\hat{\bs{\lambda}}_{M^{(s + 1)}}, \bs{\Omega}_{M^{(s + 1)}}^{-1})$ as discussed in Appendix B.\\
\item[\textbf{5.}] \textbf{Update $\bs{\Sigma}$:} Use $\bs{\lambda}^{(s + 1)}$ and $\bs{\rho}^{(s + 1)}$ to calculate $\bs{\epsilon}^{(s + 1)}$ and $\bs{\eta}^{(s + 1)}$ and sample
$$
\bs{\Sigma}^{(s + 1)} \sim \mc{W}^{-1}(\mathbb{I}_2+ \bs{Q}^{(s + 1)}, n + 3)
$$
where
$$
\bs{Q}^{(s + 1)} = [\bs{\epsilon}^{(s + 1)}\mbox{ }\bs{\eta}^{(s + 1)}]'[\bs{\epsilon}^{(s + 1)}\mbox{ }\bs{\eta}^{(s + 1)}]
$$
\end{itemize}
\noindent This constitutes the entire IVBMA algorithm. The appeal of the procedure is that it is hardly more involved than the original Gibbs sampler discussed in Section~\ref{sec:gibbs}.
\section{Empirical Analysis}
\subsection{Determinants of Macroeconomic Growth}
\begin{figure}
\subfigure[Rule of Law]{\includegraphics[height = 1.5in]{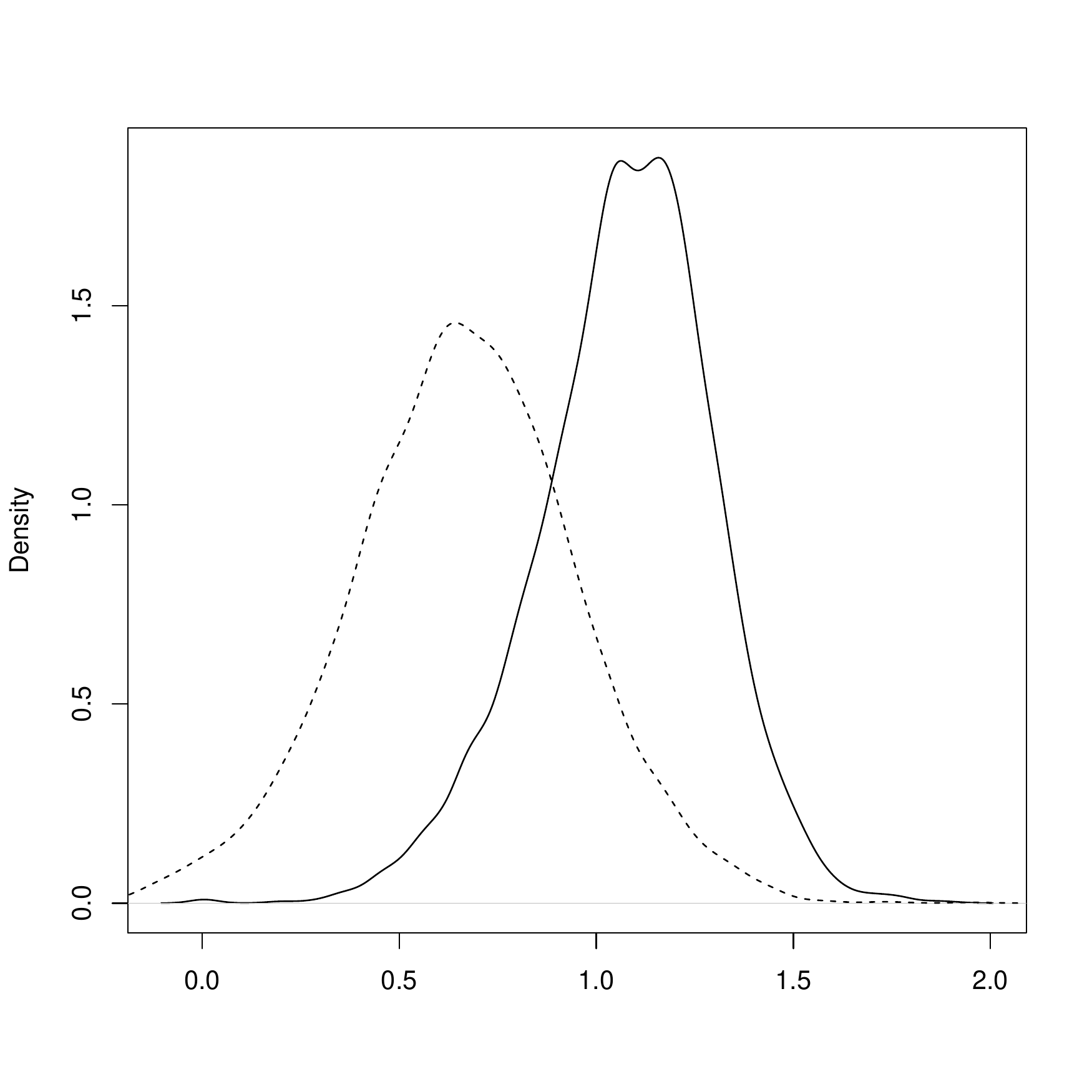}}
\subfigure[Integration]{\includegraphics[height = 1.5in]{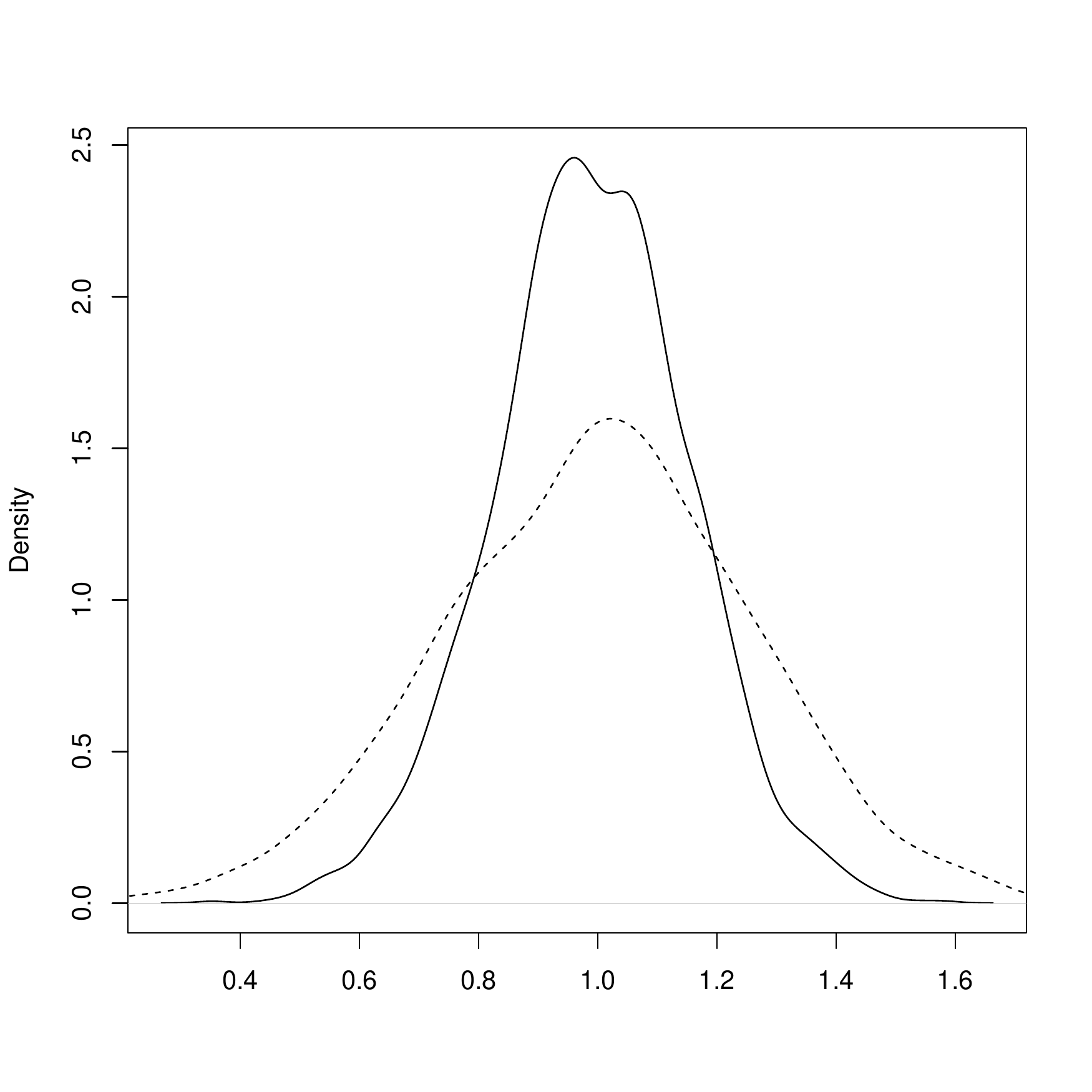}}
\subfigure[Latin America]{\includegraphics[height = 1.5in]{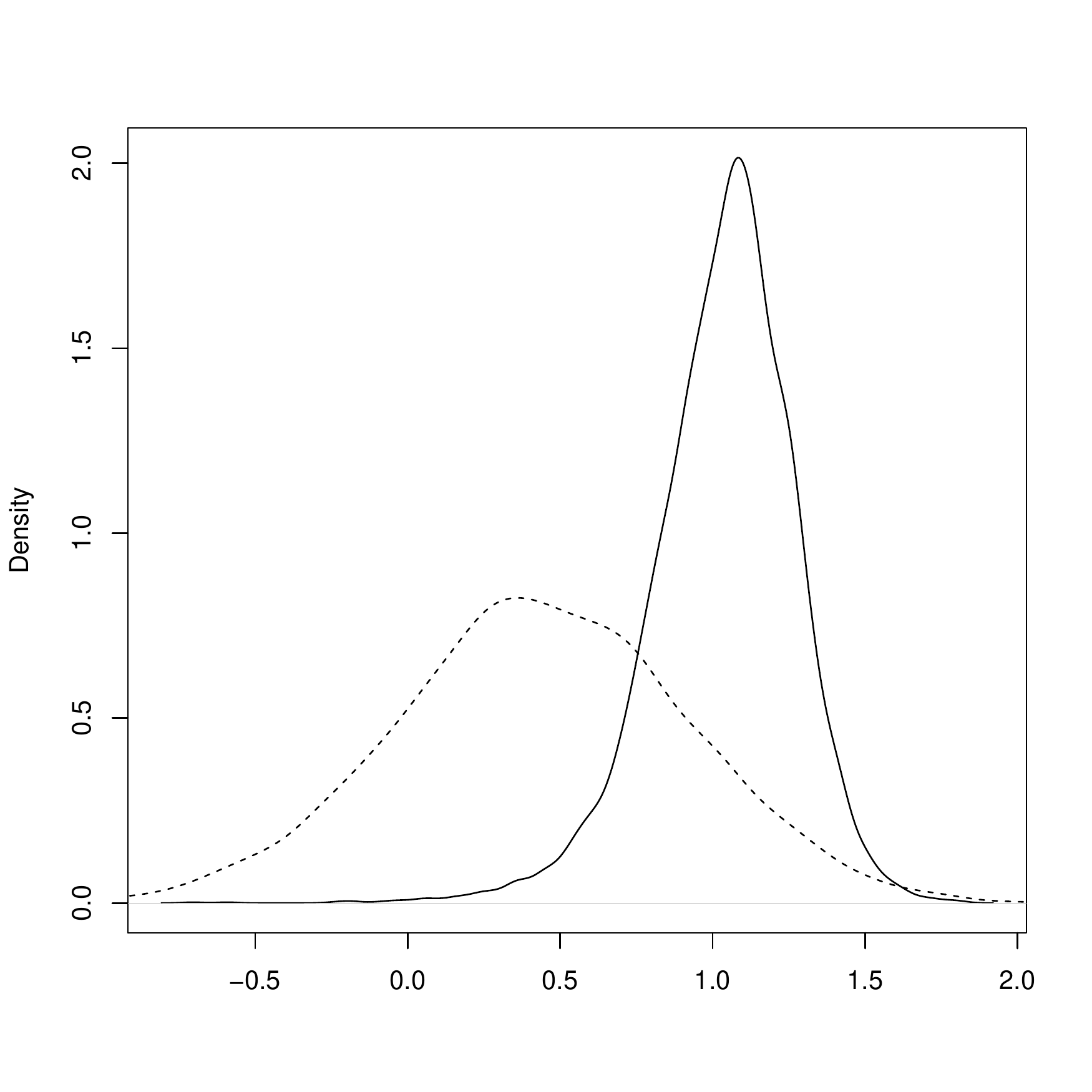}}
\subfigure[Policy Openness]{\includegraphics[height = 1.5in]{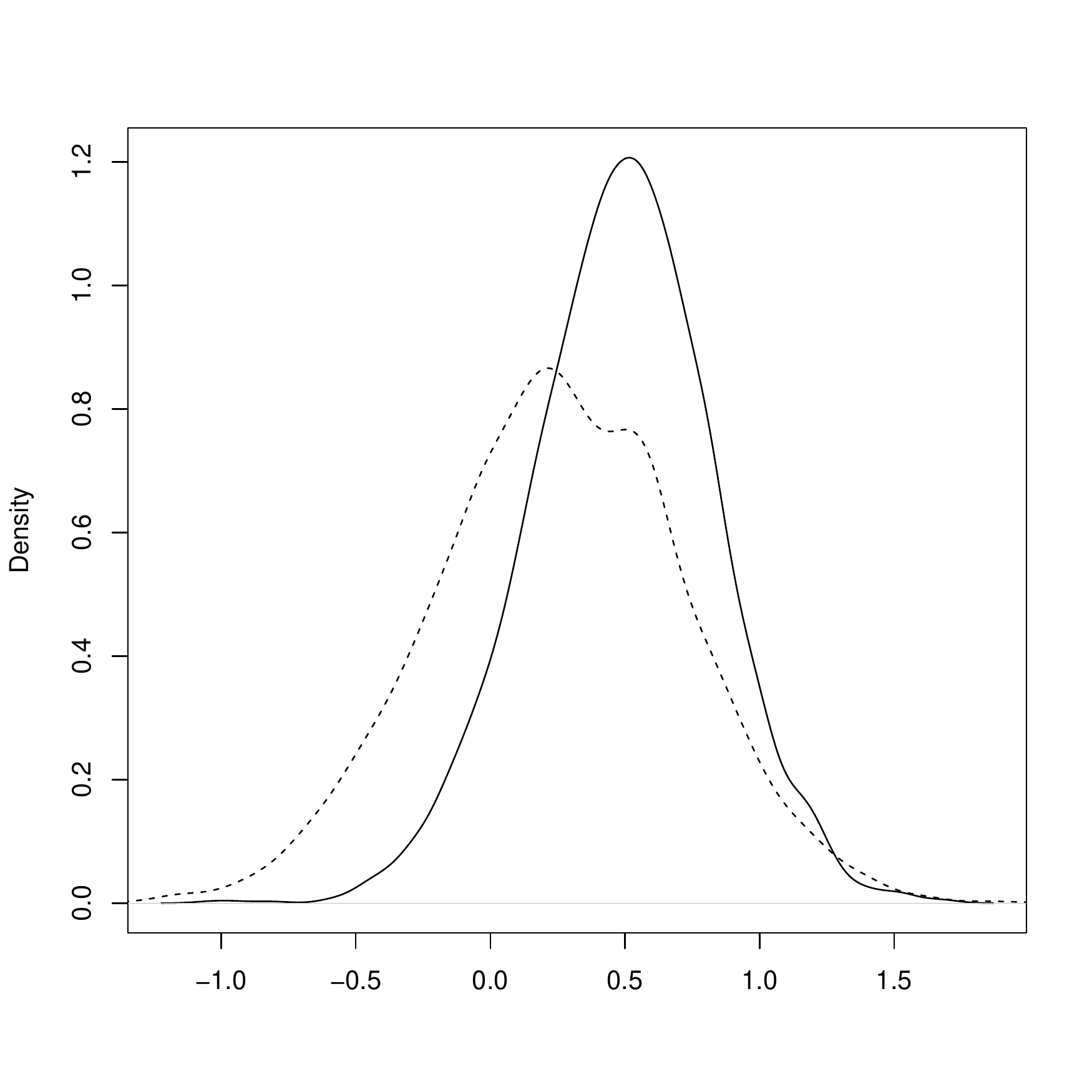}}
\caption{Posterior Distribution on selected second-stage coefficients, under IVBMA (solid line) and IV (dotted line).  In the case of IVBMA, the densities are formed conditional on inclusion in the second stage model.}\label{fig:posterior}
\end{figure}

Modeling uncertainty in macroeconomic growth determinants has proven a testing ground for BMA, see \citet{eicher_et_2007} and the extensive references therein.  We consider the dataset used in \citet{lenkoski_et_2011} which builds on that of \citet{rodrik_et_2004}.  These data juxtapose the most prominent development theories and their associated candidate regressors in one comprehensive approach. The data have two endogenous variables, a proxy for institutions (rule of law) and economic integration.  There are four potential instruments and 18 additional covariates.  Table~\ref{tab:rst} summarizes the variables included in this study.  See \citet{lenkoski_et_2011} for a detailed description of the dataset and the modeling background.\\
\linespread{1}
\begin{table}\caption{Variable Descriptions from RST dataset.}\label{tab:rst}
{\tiny
\begin{tabular}{l l}
\hline
Variable Name & Description\\
\hline
Area & Land area (thousands sq. mt.) \\
Catholic & Dummy variable taking value 1 if the country's population is predominantly catholic\\
EastAsia & Dummy variable taking value 1 if a country belongs to South-East Asia, 0 otherwise\\
EngFrac & Fraction of the population speaking English. \\
TradeShares & Natural logarithm of predicted trade shares computed from a bilateral trade equation with ``pure geography'' variables. \\
FrostArea & Proportion of land with $>$5 frost-days per month in winter. \\
FrostDays & Average number of frost-days per month in winter. \\
Integration & Natural logarithm of openness.  Openness is given by the ratio of (nominal) imports plus exports to GDP (in nominal US dollars).\\
LatinAmerica & Dummy variable taking value 1 if a country belongs to Latin America or the Caribbean, 0 otherwise\\
Latitude & Distance from Equator of capital city measured as abs(Latitude)/90\\
LegalOrigFr & variable taking a value of 1 if a country has a legal system deriving from that in France\\
LegalOrigSocialist & variable taking a value of 1 if a country has a socialist legal system\\
Malaria94 & Malaria index, year 1994. \\
MeanTemp & Average temperature (Celsius). \\
Muslim & Dummy variable taking value 1 if the country's population is predominantly muslim\\
Oil & variable taking value 1 for a country being major oil exporter, 0 otherwise.  \\
PolicyOpenness & Dummy variable that indicates if a country has sufficiently market oriented policies\\
PopGrowth & population growth\\
Protestant & variable taking value 1 if the country's population is predominantly protestant\\
RuleofLaw & Rule of Law index. Refers to 2001 and approximates for 1990's institutions \\
SeaAccess & Dummy variable taking value 1 for countries without access to the sea, 0 otherwise.  \\
SettlerMortality & Natural logarithm of estimated European settlers' mortality rate\\
SubSaharaAfrica & taking value 1 if a country belongs to Sub-Saharan Africa, 0 otherwise\\
Tropics & Percentage of tropical land area.\\
\hline
\end{tabular}
}
\end{table}
\linespread{1.3}
\indent We took the dataset of \citet{lenkoski_et_2011} and ran IVBMA for 200,000 iterations, discarding the first 20,000 as burn-in.  This took approximately 10 minutes to run.  By contrast, the 2SBMA analysis conducted by \citet{lenkoski_et_2011} on the same data took over 15 hours of computing time.  The extreme difference in computing time results from the style of the two approaches.  The 2SBMA methodology of \citet{lenkoski_et_2011} was designed to mimic the 2SLS estimator.  It first ran a separate BMA analysis for each first-stage regression.  All models returned from these two runs were paired and a subsequent BMA was run on the outcome equation for each pair.  This led to an extremely large number of second-stage BMA runs and thus considerable computing time.  By contrast, IVBMA models the entire system jointly and this joint approach leads to a dramatic improvement in computational efficiency.\\
\indent Table~\ref{tab:rst_results} shows the resulting posterior estimates.  We see a picture similar to that reported by \citet{lenkoski_et_2011}, although with somewhat fewer included determinants. In particular, similar to \citet{lenkoski_et_2011} English and European fractions serve as the two best instruments of Rule of Law, while neither settler mortality nor trade receive high inclusion probabilities.  Further, Integration is well-instrumented by trade shares, which receives an inclusion probability of $1$.  These results are essentially the same as those reported in \citet{lenkoski_et_2011}.\\
\indent In the second stage, we see a similar, but markedly sparser conclusion as \citet{lenkoski_et_2011}.  Both rule of law and integration are given strong support by the data, with inclusion probabilities of essentially $1$.  Beyond these two factors only the intercept, an indicator for Latin America and an indicator of whether the country has market oriented policies are given inclusion probabilities above 0.5 in the second stage.  In contrast to 2SBMA, which gave evidence to religious and geographic issues as determinants of macroeconomic growth, IVBMA points strongly to institutions and integration as the leading determinants.\\
\linespread{1}
\begin{table}\caption{Results for Macroeconomic Growth Determinants Example.  In this table we show the posterior inclusion probabilities (Prob), posterior parameter expectations (Mean) and posterior standard deviations (sd) for the two instrument stages as well as the outcome stage.}\label{tab:rst_results}
\begin{center}
{\tiny
\begin{tabular}{lccccccccc}
\hline
\hline
 & \multicolumn{3}{c}{Rule}&\multicolumn{3}{c}{Trade}&\multicolumn{3}{c}{Outcome}\\
Variable & Prob & Mean & sd & Prob & Mean & sd & Prob & Mean & sd\\
\hline
RuleofLaw&--&--&--&--&--&--&0.999&1.073&0.224\\
Integration&--&--&--&--&--&--&1&0.992&0.164\\
SettlerMortality&0.11&-0.009&0.035&0.097&-0.006&0.028&--&--&--\\
TradeShares&0.111&0.007&0.037&1&0.532&0.088&--&--&--\\
EnglishFrac&0.91&1.13&0.592&0.539&0.244&0.302&--&--&--\\
EuropeanFrac&0.667&0.459&0.455&0.16&-0.012&0.087&--&--&--\\
Intercept&0.271&0.061&0.278&0.999&2.303&0.343&0.546&0.362&0.793\\
Dist\_Equ&0.016&0&0.002&0.007&0&0.001&0.015&0&0.002\\
Lat\_Am&0.539&-0.297&0.371&0.163&-0.017&0.077&0.981&1.018&0.271\\
Sub\_Africa&0.207&-0.029&0.114&0.184&0.027&0.099&0.233&-0.034&0.133\\
E\_Asia&0.415&0.152&0.269&0.957&0.671&0.261&0.381&0.126&0.288\\
Legor\_fr&0.157&0.008&0.067&0.122&0.008&0.045&0.366&0.101&0.173\\
Catholic&0.007&0&0.001&0.003&0&0&0.031&0&0.002\\
Muslim&0.002&0&0&0.002&0&0&0.017&0&0.001\\
Protestant&0.008&0&0.001&0.021&0&0.001&0.011&0&0.001\\
Tropics&0.939&-0.607&0.24&0.338&0.095&0.177&0.381&-0.131&0.263\\
SeaAccess&0.165&-0.012&0.075&0.119&0.001&0.046&0.158&-0.005&0.085\\
Oil&0.344&-0.108&0.212&0.881&0.402&0.21&0.387&0.145&0.266\\
Frost\_Day&0.04&0.001&0.006&0.022&0&0.002&0.03&0&0.005\\
Frost\_Area&0.465&0.194&0.289&0.376&0.133&0.24&0.341&0.071&0.253\\
Malaria94&0.242&-0.042&0.136&0.301&-0.076&0.152&0.243&-0.035&0.149\\
MeanTemp&0.021&0&0.004&0.017&0&0.002&0.026&0&0.005\\
Area&0&0&0&0&0&0&0&0&0\\
Population&0.037&-0.001&0.011&0.042&0&0.009&1&0.235&0.04\\
PolicyOpen&0.37&0.118&0.236&0.228&0.021&0.132&0.511&0.249&0.345\\
\hline
\end{tabular}
}
\end{center}
\end{table}
\indent Figure~\ref{fig:posterior} shows the posterior distribution of the second-stage coefficients for the four variables with the highest inclusion probabilities under IVBMA.  We also include the posterior distribution of these covariates under an approach that does not incorporate model uncertainty (which we refer to as IV), and uses the algorithm discussed in Section~\ref{sec:gibbs}.  Several interesting aspects are clear in Figure~\ref{fig:posterior}.  Inspecting panel (b), we see that IVBMA has led to a posterior distribution on integration with essentially the same mode as that of IV.  However, the IVBMA distribution is considerably more focused, indicating a reduction in parameter variance that results from using parsimonious models.\\
\indent The other three panels also have the feature of tighter posterior distributions under IVBMA versus IV.  However, what is potentially more interesting is that the distributions are also centered in slightly different locations.  The effect is particularly large for the Latin America indicator, which is tightly centered about its median of $1.06$ under IVBMA, while more diffuse about the median of $0.43$ under IV.  The respective posterior standard deviations of these two estimates are $0.233$ and $0.486$ under IVBMA and IV respectively.\\
\indent This effect is also evident for the rule of law parameter estimate.  Under IVBMA, this parameter has a median of $1.08$ and posterior standard deviation of $0.224$, while under IV this parameter has a median of $0.666$ and an standard deviation of $0.284$.  We note that in \citet{lenkoski_et_2011}, three increasingly larger runs of 2SBMA were conducted.  As the size of the considered covariates rose, the posterior estimate on rule of law went from 1.276 (with a standard deviation 0.1772) down to an estimate of $.798$ ($.3155$).  Therefore, our results are in line with those of \citet{lenkoski_et_2011}, however it appears evident that IVBMA has introduced additional parsimony into resulting models than the nested approach of 2SBMA.
\linespread{1.3}
\subsection{Estimating a Demand Function}\label{sec:margine}
\subsubsection{Description of the Data Set}
We use the data provided by \citet{chintagunta_et_2005} (CDG) that had been collected by AC Nielsen and follow the approach outlined in \citet{conley_hansen_rossi}. CDG examined the purchase of margarine in Denver, Colorado, in a time period of 117 weeks, from January 1993 until March 1995. The sample consists of weekly prices and purchase data for the four main brands of margarine. CDG differentiate between 992 households purchasing margarine whereas following \citet{conley_hansen_rossi} we will not account for heterogeneity but focus on the total number of weekly purchases per brand. Furthermore, the data set offers weekly information on feature ads and display conditions for each of the four brands. For detailed descriptive statistics and marketing conditions of the single brands see \citet{chintagunta_et_2005}.\\
\indent Since retail price is influenced by unobserved characteristics likely to be correlated with sales, it is an endogenous variable. CDG claim that wholesale prices serve as reliable instruments as they should not be sensitive to retail demand shocks. Their results show that wholesale prices alone explain nearly 80\% of the variation in margarine retail prices. Moreover, it is often the case that products with considerable shelf-life such as magarine are not sold to the consumer within the same week as they are bought at the wholesale establishment. Thus, CDG added the wholesale prices of up to six weeks before the purchase week to the matrix of instruments.\\
\indent Besides these variables we entertain two more candidate instruments. We include the Consumer Price Index (CPI) for all urban consumers of Colorado and the CPI for food in the United States, using the data provided by the U.S. Bureau of Labor Statistics (BLS). Since the BLS reports only monthly data, we use the same value for all weeks in the respective month. Weeks being part of two months are assigned to the month the majority of their days belong to. We do not expect these variables to perform as well as wholesale prices because they are not collected at a brand level. However, we think it is reasonable (or at least vaguely plausible) that overall price levels should influence the price of margarine. Our matrix $\boldsymbol{Z}$ therefore consists of nine candidate instrumental variables (see table \ref{variables} for an overview).\\
\indent In addition to feature ads and display conditions, we entertain several additional variables with potential effect on both demand and retail price. Our hypothesis is that holidays could positively affect the demand for margarine. We therefore collected data from the Denver Public Schools showing the days free of school for the school years 1992/93, 1993/94 and 1994/95. Differing between whole weeks of holiday and weeks containing only one or two free days, we created two dummy variables and added them to the matrix $\bs{W}$.\\
\indent We also consider the Local Area Unemployment Statistics (LAUS) of Colorado. These monthly data provided by the BLS are again adapted to our weekly setup in the manner described above.\\
\indent Moreover, we entertain the possibility that temperature might also have explanatory power for the purchase of margarine.  We therefore collected historical temperature data for the Denver area from January 1993 until March 1995. Finally, we add four fixed variables to $\boldsymbol{W}$ for distinguishing between brands. Table \ref{variables} summarizes the different regressors by short descriptions.\\
\indent Following \citet{conley_hansen_rossi}, we examine the logarithm of each brand's weekly share of sales instead of the absolute sales figures. Additionally, we use the logarithm of retail prices as endogenous regressors, yielding the regression system
\begin{align*}
\text{log }Share&= \beta\text{ log}(retail\;price)+\boldsymbol{W}\boldsymbol{\gamma}+\epsilon \\
\text{log}(retail\;price)&=\boldsymbol{Z}\boldsymbol{\delta}+\boldsymbol{W}\boldsymbol{\tau}+\eta. 
\end{align*}
These transformations are clearly performed in order to use the framework (\ref{equ1}) and (\ref{equ2}).   A more involved specification would directly assess the discrete choice nature of the dataset; we discuss this feature in the Conclusions section.
\linespread{1}
\begin{table}
{\footnotesize
\begin{tabular}{l l}
\hline
Variable Name & Description\\
\hline
WP & weekly wholesale prices for the four different brands for margarine\\
lag1 WP & wholesale prices one week before sale to consumer\\
lag2 WP & wholesale prices two weeks before sale to consumer \\
lag3 WP & wholesale prices three weeks before sale to consumer\\
lag4 WP & wholesale prices four weeks before sale to consumer \\
lag5 WP & wholesale prices five weeks before sale to consumer \\
lag6 WP & wholesale prices six weeks before sale to consumer\\
CPI Food & CPI for food in general in the U.S.\\
CPI UrbCol & CPI for all urban consumers of Colorado\\
\hline
Feature Ad & variable indicating the existance and degree of feature ads at the product shelfs\\
Display & variable describing the display conditions\\
Intercept & vector with value 1, reference point for brand indicators \\
Brand2 & dummy variable indicating brand 2\\
Brand3 & dummy variable indicating brand 3\\
Brand4 & dummy variable indicating brand 4\\
WeekHol & dummy variable taking value 1 if the whole week was free at Denver Public Schools\\
InterHol & dummy variable taking value 1 if the week had only one or two free days at DPSs\\
Temp & variable showing the average weekly temperature at Denver, Colorado (in Celsius)\\
Unemploy & Local Area Unemployment Statistics for Colorado \\
\hline
\end{tabular}
}
\caption{Descriptions of the variables contained in $\boldsymbol{Z}$ and $\boldsymbol{W}$ (upper and lower part of the table, respectively).} \label{variables}
\end{table}
\subsubsection{Results - Factors influencing the Demand for Margarine} 
\begin{figure}
\begin{center}
\subfigure[Instrument Stage]{\includegraphics[height=3in]{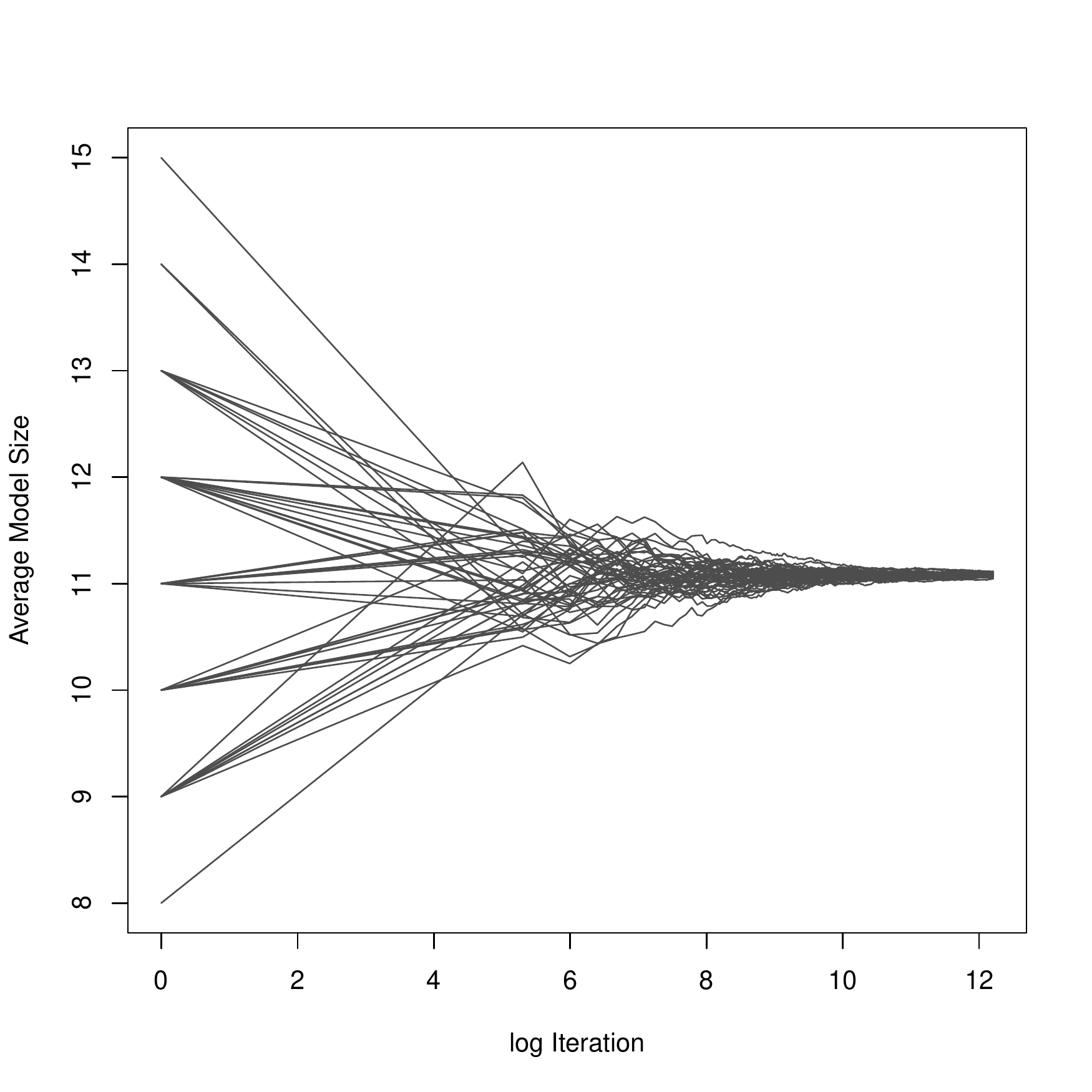}}
\subfigure[Outcome Stage]{\includegraphics[height=3in]{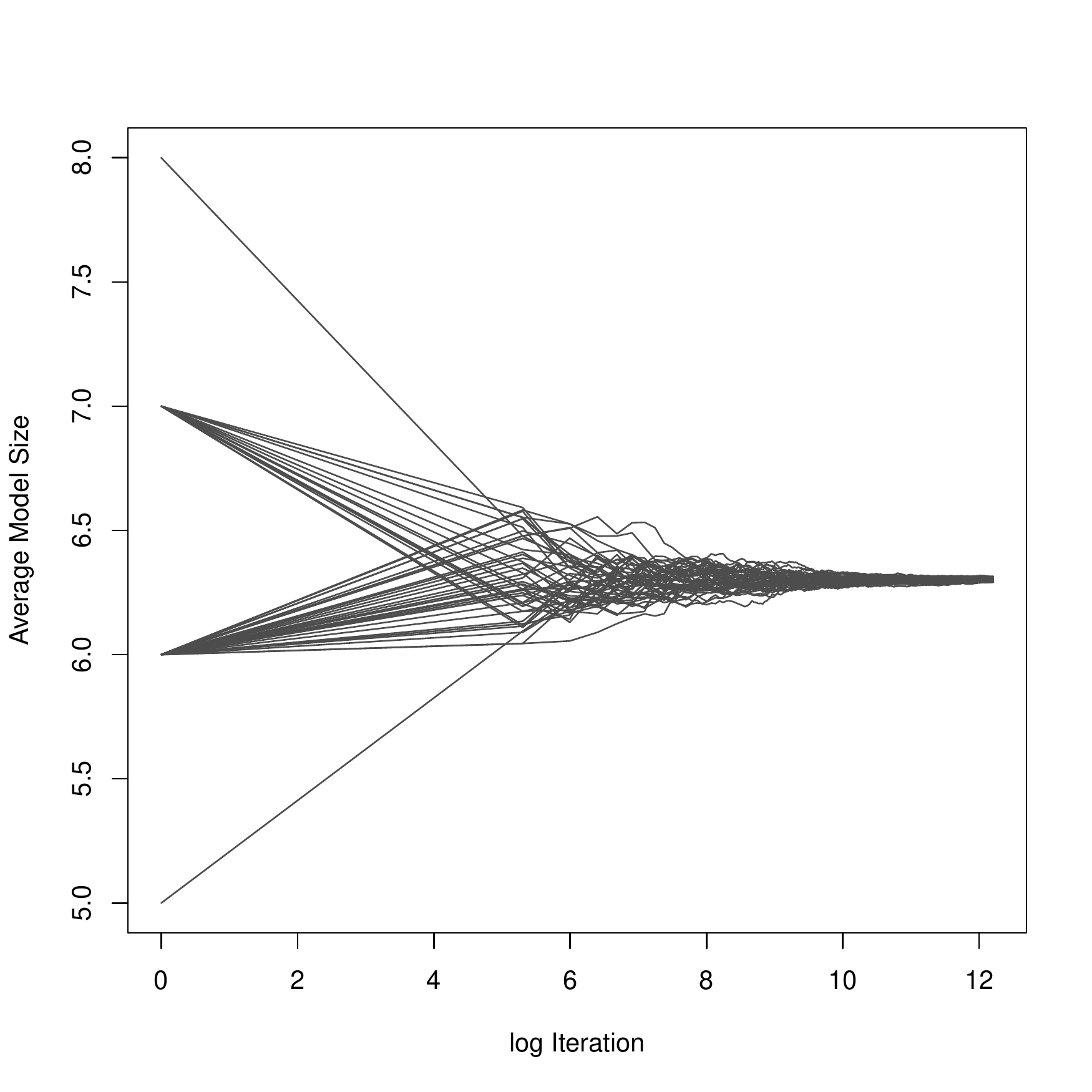}}
\end{center}
\caption{Average model size by log iteration for the instrument and outcome models across $50$ separate instances of IVBMA.  This provides a rough diagnostic that the IVBMA chains have converged.}\label{fig:converge}
\end{figure}

\indent For the margarine data we considered 19 potential influencing factors in the first stage, amongst them 9 instruments. In the second stage, we chose 10 variables to predict the log shares of sales.\\
\indent We ran IVBMA for 250,000 iterations and discarded the first 50,000 as burn-in.  In order to examine the mixing properties of IVBMA, we ran 50 independent instances of the algorithm initialized at different random starting points and using different random seeds.  On average, each run took approximately 5 minutes on the hardware discussed above and all $50$ instances returned identical posterior estimates, indicating convergence and no issues regarding mixing.  Figure~\ref{fig:converge} shows a rough diagnostic of this convergence.  In it, we show the average first stage (Equation \ref{equ2}) and second stage (Equation \ref{equ1}) model size by log iteration for each of the $50$ chains.  As we can see, the figure shows a rapid agreement across chains, with an average model size of $11.08$ in the first (instrument) stage and $6.30$ in the second (outcome) stage.  While this visual display is only a rough diagnostic, it gives an idea of the quick convergence and lack of mixing difficulties of IVBMA. Indeed, it appears that $250,000$ iterations may have been unnecessary as all chains agree within the first $50,000$ post burn-in iterations.\\
\indent Columns 1 and 5 of Table \ref{parest} show the inclusion probabilities of both stages returned by IVBMA. We note that the log price, the endogenous factor, is given an inclusion probability of 1. The other columns in Table \ref{parest} provide the lower, median, and upper bounds (the 2.5\%, 50\% and 97.5\% quantiles respectively) of the resulting parameter samples. The posterior distribution of $\beta$ has a median of -2.161 with a small range, confirming the expectation that higher retail price of margarine diminishes quantity sold.\\
\indent Regarding the results for members of $\boldsymbol{W}$, we see that in both stages the feature ad and display variables have inclusion probabilities of more than 95\%. The median parameter values in Table \ref{parest}, columns 3 and 6, indicate that feature ads and display conditions have a negative effect on price but simultaneously a positive influence on the volume of demand.\\
\indent With regard to variables we have added, temperature proves to affect neither prices nor sales figures. In our mind this shows the utility of IVBMA, as we were able to entertain this additional factor and the method promptly rejected its inclusion. Similar results are found for both holiday variables, which have an inclusion probability of less than 6\% and therefore little influence in either stage.\\
\indent However, the unemployment rate for Colorado offers an unexpected surprise.  This variable has an impressively high inclusion probability of 87\% in the first stage and a negligible influence at the second stage.  Therefore, it seems that unemployment could serve as an instrument for the endogenous price variable.  We feel this could be reasonable, largely because our dependent variable is share and not quantity sold.  Thus, we can imagine declining economic conditions to induce retailers to lower the price of consumer staples, but these conditions may have potentially less effect on the overall product mix sold once price adjustments are accounted for.  As we can see in column 2 of Table \ref{parest} this factor is a negatively directed factor, i.e. the higher the unemployment rate in Colorado the lower the price for margarine.\\
\indent Use of wholesale prices as instruments is confirmed, but not wholeheartedly. These variables' inclusion probabilities ranged between 48\% and 66\% (column 1, Table \ref{parest}). Interestingly, the effect of wholesale prices increases with the weekly time-lags, which is also reflected in the posterior medians of the regression coefficients in column 3 of Table \ref{parest}. As already reasoned above, this could be grounded in the fact retailers often buy their products from the wholesalers some time before they are sold to the consumer. Besides wholesale prices, the CPI variables also seem to have some, albeit limited, influence on the retail price of margarine. The CPI for food and the CPI of urban consumers are given inclusions probabilities of 30\% and 13\%, respectively.\\
\linespread{1}
\begin{table}
{\footnotesize\linespread{1.4}
\begin{center}
\begin{tabular}{l ccccc cccc}
\hline
 & \multicolumn{4}{c}{First Stage}&\multicolumn{4}{c}{Second Stage}\\
Variable & Prob & Lower & Median & Upper & Prob & Lower & Median & Upper\\
\hline
log price & -- & -- & -- & -- &1 & -2.839 & -2.161 & -1.637\\
feat.W1 & 0.965 & -0.069 & -0.046 & 0 & 0.995 & 0.182 & 0.375 & 0.559\\
disp.W2 & 0.97 & -0.306 & -0.195 & 0 & 0.997 & 0.678 & 1.508 & 2.311\\
Int & 0.999 & -2.603 & -1.417 & -1.261 & 1 & -5.508 & -4.556 & -3.86\\
Brand2 & 1 & 0.245 & 0.338 & 0.426 & 1 & 0.831 & 1.031 & 1.33\\
Brand3 & 1 & 0.119 & 0.141 & 0.163 & 0.997 & 0.152 & 0.267 & 0.408\\
Brand4 & 1 & 0.172 & 0.198 & 0.222 & 0.137 & -0.014 & 0 & 0.19\\
InterHol & 0.006 & 0 & 0 & 0 & 0.057 & -0.009 & 0 & 0.009\\
WeekHol & 0.012 & 0 & 0 & 0 & 0.059 & 0 & 0 & 0.044\\
Temp & 0 & 0 & 0 & 0 & 0.007 & 0 & 0 & 0\\
Unemploy & 0.869 & -0.031 & -0.022 & 0 & 0.054 & -0.036 & 0 & 0\\
lag6.Z6 & 0.654 & -0.503 & 0.447 & 2.68& -- & -- & -- & -- \\
lag5.Z5 & 0.603 & -0.657 & 0.047 & 2.487& -- & -- & -- & -- \\
lag4.Z4 & 0.548 & -0.842 & 0 & 2.246& -- & -- & -- & -- \\
lag3.Z3 & 0.529 & -0.926 & 0 & 2.115& -- & -- & -- & -- \\
lag2.Z2 & 0.509 & -1.073 & 0 & 1.94& -- & -- & -- & -- \\
lag1.Z1 & 0.485 & -1.193 & 0 & 1.765& -- & -- & -- & -- \\
WP.Z0 & 0.5 & -1.097 & 0 & 1.841& -- & -- & -- & -- \\
CPIFood & 0.302 & -0.682 & 0 & 0.503& -- & -- & -- & -- \\
CPIUrb & 0.135 & 0 & 0 & 0.008& -- & -- & -- & -- \\
\hline
\end{tabular}
\end{center}
}
\caption{IVBMA results for the margarine dataset.  This table shows first and second stage inclusion probabilities (Prob) as well as 2.5\%, 50\% and 97.5\% posterior quantiles (Lower, Median, Upper respectively) for each variable included.}\label{parest}
\end{table}
\section{Conclusion}
\indent We have proposed a computationally efficient solution to the problem of incorporating model uncertainty into IV estimation.  The IVBMA method leverages an existing Gibbs sampler and shows that by nesting model moves inside this framework, model averaging can be performed with minimal additional effort.  In contrast to the approximate solution proposed by \citet{lenkoski_et_2011}, our method yields a theoretically justified, fully Bayesian procedure. The applied examples shows the utility the method offers, by enabling additional factors to be entertained by the researcher, which are either incorporated where appropriate or promptly dropped.\\
\indent The RJMCMC methodology proposed by \citet{koop_et_2012} constitutes an alternative approach to this problem.  Their method is considerably more flexible; it allows a range of different prior distributions to be entertained and simultaneously addresses hypotheses related to identification in the IV system.  At the same time, this flexibility comes at a cost.  \citet{koop_et_2012} note that their method may exhibit difficulties in mixing and are required to consider a complicated model proposal system involving ``hot'',``cold'', and ``super-hot'' models which has similarities to simulated tempering.  In contrast IVBMA appears to exhibit few difficulties in mixing, which derives from the simplicity of the algorithm.  We feel that, at the very least, IVBMA offers a useful methodology for the applied researcher, who may be willing to accept the priors we propose in order to quickly obtain useful insight and parameter estimates. \\
\indent In the IV framework we develop, we consider only one endogenous variable for clarity of exposition.  Multiple endogenous variables pose no significant additional difficulties.  The Gibbs sampler in Section \ref{sec:gibbs} requires repeated evaluations of a slightly modified Step 2.  The IVBMA framework simply consists of different first-stage models $M$ for each endogenous variable. The CBFs are hardly changed. This generalization of our framework has already been incorporated into the {\tt R} package {\tt ivbma}.\\
\indent One assumption that is crucial to the functioning of the Gibbs sampler is the bivariate normality of the residuals in (\ref{err}).  \citet{conley_et_2008} discuss how the algorithm of \citet{rossi_et_2006} can be extended to handle deviations from normality using a Dirichlet process mixture (DPM).  We note that the IVBMA methodology can readily be incorporated into the DPM framework of \citet{conley_et_2008} simply by replacing the IV kernel distributions of \citet{rossi_et_2006} with IVBMA kernel distributions.\\
\indent A critical feature that has not been addressed in IVBMA is that of instrument validity.  \citet{lenkoski_et_2011} propose an approximate test of instrument validity by directly embedding the test of \citet{sargan_1958} into a model averaging framework.  While this appears to work well, we are currently researching a ``fully-Bayesian'' version of the Sargan test which is based on the CBF of regressing the instrument set on the residuals of (\ref{equ1}).  Subsequent research will develop this test and incorporate it into the IVBMA method. A proto-type of this diagnostic is already implemented in {\tt ivbma}.\\
\indent Finally, as we note, the margarine dataset is a simplification, as it ignores the aspect of multinomial choice and significantly reduces the household information collected.  Following \citet{conley_hansen_rossi}, log shares were used to fit the data into the framework (\ref{equ1}) and (\ref{equ2}).  However, we feel that IVBMA has the potential to be extended to more complicated likelihood frameworks.  Since discrete choice models may be represented in a generalized linear model (GLM) framework with latent Gaussian factors (for instance via a multinomial probit), a promising next step will be to consider embedding IVBMA in a GLM model and operating on these latent factors.  In our mind, this indicates the true potential benefit of IVBMA.  Since the entire method uses a Gibbs framework, it may be incorporated in any setting where endogeneity, model uncertainty and latent Gaussianity are present.
\section{Acknowledgements}
The authors would like to thank Pradeep Chintagunta for supplying the margarine dataset, Theo S. Eicher for several helpful comments and Andreas Neudecker for support organizing the software associated with this work.  Alex Lenkoski gratefully acknowledges support from the German Science Foundation (DFG), grant GRK 1653.\\
\bibliographystyle{chicago}
\bibliography{KarlLenkoski}
\renewcommand{\theequation}{A-\arabic{equation}} 
\setcounter{equation}{0}  
\section*{Appendix A}
\textbf{Details of the determination of $pr(\bs{\rho}|\bs{\lambda}, \bs{\Sigma}, \mc{D})$}\\
\indent Our derivation follows \citet{rossi_et_2006} closely--extended to the multivariate setting.  Set $\bs{V} = [X\mbox{ } \bs{W}]$.  Then, conditional on $\bs{\Sigma}$ and $\bs{\lambda}$ we have
\begin{align*}
\bs{Y}&=\bs{V}\bs{\rho} + \bs{\epsilon}\\
&=\bs{V}\bs{\rho} + \frac{\sigma_{21}}{\sigma_{22}}\bs{\eta}+\bs{\nu}_{1|2},\label{step1Y}
\end{align*}
where $\bs{\eta}$ is derived from $\bs{\lambda}$ and $\mc{D}$ and $(\bs{\nu}_{1|2})_i\thicksim\mathcal{N}\left(0,\xi \right)$ where $\xi = \sigma_{11} - \sigma_{12}^2/\sigma_{22}$.\\
Replacing $\bs{Y}$ by $\tilde{\bs{Y}}= \bs{Y} - (\sigma_{21}/\sigma_{22})\bs{\eta}$ yields.
$$
\tilde{\bs{Y}}=\textbf{\textit{V}}\boldsymbol{\rho}+\bs{\nu}_{1|2}.
$$
We now compute
\begin{align*}
pr(\bs{\rho}|\tilde{\bs{Y}},\bs{V})&\propto pr(\tilde{\bs{Y}}|\bs{\rho},\bs{V}) pr(\bs{\rho})\\
&\propto \exp\left(-\frac{1}{2\xi} (\tilde{\bs{Y}}-\bs{V}\bs{\rho})' (\tilde{\bs{Y}}-\bs{V}\bs{\rho})-\frac{1}{2}\bs{\rho}'\bs{\rho}\right)\\
&\propto \exp\left(-\frac{1}{2}\left[-2\xi^{-1}\tilde{\bs{Y}}'\bs{V}\bs{\rho}+\bs{\rho}'(\mathbb{I}_{1+p}+\xi^{-1}\bs{V}'\bs{V})\bs{\rho} \right]\right).
\end{align*}
Setting $\bs{\Xi} = \mathbb{I}_{1+p}+\xi^{-1}\bs{V}'\bs{V}$ and $\hat{\bs{\rho}} = \xi^{-1}\tilde{\bs{Y}}'\bs{V}\bs{\Xi}^{-1}$, this becomes
\begin{align*}
pr(\bs{\rho}|\tilde{\bs{Y}},\bs{V}) &\propto\exp\left(-\frac{1}{2}\left[-2\hat{\bs{\rho}}'\bs{\Xi}\bs{\rho} + \bs{\rho}\bs{\Xi}\bs{\rho}\right]\right)\\
&\propto\frac{|\bs{\Xi}|^{1/2}}{(2\pi)^{(1 + p)/2}}\exp\left(-\frac{1}{2}(\bs{\rho} - \hat{\bs{\rho}})'\bs{\Xi}(\bs{\rho} - \hat{\bs{\rho}})\right).
\end{align*}
Thus, we conclude
$$
\boldsymbol{\rho}|\boldsymbol{\lambda},\boldsymbol{\Sigma},\mathcal{D} \thicksim\mathcal{N}(\hat{\bs{\rho}},\bs{\Xi}^{-1}),
$$
which confirms \eqref{distrho}.\\
\noindent \textbf{Details of the determination of $pr(\bs{\lambda}|\bs{\rho}, \Sigma, \mc{D})$}\\
\indent We now provide a detailed derivation of \eqref{distlambda}.\\
Inserting \eqref{equ2} into \eqref{equ1} leads to
$$
\bs{Y}=\bs{Z}\bs{\delta}\beta+\bs{W}\bs{\tau}\beta+\boldsymbol{W}\boldsymbol{\gamma}+\beta\bs{\eta}+\bs{\epsilon}.
$$
Conditioning on $\beta$ and $\boldsymbol{\gamma}$ we set $\bs{Y}^*=\beta^{-1}(\bs{Y}-\boldsymbol{W}\boldsymbol{\gamma})$ and obtain $\bs{Y}^*=\boldsymbol{Z}\boldsymbol{\delta}+\boldsymbol{W}\boldsymbol{\tau}+\bs{\vartheta}$ with $\bs{\vartheta}=\bs{\eta}+\beta^{-1}\bs{\epsilon}$.  Further $\vartheta_i\thicksim\mathcal{N}(0,\zeta)$, with $\zeta=\sigma_{22}+\beta^{-2}\sigma_{11}+2\beta^{-1}\sigma_{21}$.\\
We can now write this as a regression system in which the number of observations has been doubled to $2n$,
\begin{equation}\label{doubled}\begin{pmatrix}\bs{Y}^* \\ \bs{X}\end{pmatrix}=\begin{pmatrix}\boldsymbol{Z} \\ \boldsymbol{Z}\end{pmatrix}\boldsymbol{\delta}+\begin{pmatrix}\boldsymbol{W} \\ \boldsymbol{W}\end{pmatrix}\boldsymbol{\tau}+\begin{pmatrix}\bs{\vartheta} \\ \bs{\eta}\end{pmatrix} \end{equation}
with  
\begin{equation*} \begin{pmatrix}\vartheta_i \\ \eta_i\end{pmatrix}\thicksim\mathcal{N}_{2}(0,\boldsymbol{\Psi})\quad\text{and}\quad \boldsymbol{\Psi}=\begin{pmatrix}\sigma_{22}+\frac{1}{\beta^2}\sigma_{11}+\frac{2}{\beta}\sigma_{21} &\sigma_{22}+\frac{1}{\beta}\sigma_{12} \\ \sigma_{22}+\frac{1}{\beta}\sigma_{12} &\sigma_{22}\end{pmatrix}.\end{equation*}
Let $\bs{\Phi}$ be the Cholesky decomposition of $\bs{\Psi}$. We then post-multiply two copies of each component in Equation \eqref{doubled} by $\boldsymbol{\Phi}^{-1}$, to obtain a regression system with unit covariance matrix for the error terms.\\
Let 
\begin{align*}
 [\hat{\bs{Y}} \enspace \hat{\bs{X}}]&=[\bs{Y}^* \enspace \bs{X}]\boldsymbol{\Phi}^{-1},\\
[\hat{\boldsymbol{Z}}_j \enspace \hat{\boldsymbol{\mathcal{Z}}}_j]&=[\bs{Z}_{j} \mbox{ } \bs{Z}_{j}]\boldsymbol{\Phi}^{-1},\mbox{ }j=1,\dots,q\\
[\hat{\boldsymbol{W}_k} \enspace \hat{\boldsymbol{\mathcal{W}}}_k]&=[\bs{W}_{k} \mbox{ } \bs{W}_{k}]\bs{\Phi}^{-1}\mbox{ }k=1,\dots,p \\
[\hat{\bs{\vartheta}} \enspace \hat{\bs{\eta}}]&=[\bs{\vartheta} \enspace \bs{\eta}]\boldsymbol{\Phi}^{-1}.
\end{align*}
This yields
\begin{align*} 
  \begin{pmatrix}\hat{\bs{Y}}\\ \hat{\bs{X}}\end{pmatrix}&=\begin{pmatrix}\hat{\boldsymbol{Z}} & \hat{\boldsymbol{W}}\\ \hat{\boldsymbol{\mathcal{Z}}} & \hat{\boldsymbol{\mathcal{W}}}\end{pmatrix}\boldsymbol{\lambda}+\begin{pmatrix}\hat{\bs{\vartheta}}\\ \hat{\bs{\eta}}\end{pmatrix} \quad\text{with }\begin{pmatrix}\hat{\vartheta}_i\\ \hat{\eta}_i\end{pmatrix}\thicksim\mathcal{N}(0,\mathbb{I}_2).
\end{align*}
Set $\boldsymbol{S}=[\hat{\bs{Y}}'\mbox{ }\hat{\bs{X}}']'$ and $\boldsymbol{T}=\begin{pmatrix}\boldsymbol{Z} & \hat{\boldsymbol{W}}\\ \hat{\boldsymbol{\mathcal{Z}}} & \hat{\boldsymbol{\mathcal{W}}}\end{pmatrix}$. The posterior distribution of $\boldsymbol{\lambda}$ is determined by the same logic as in Step 1 and gives
$$
\boldsymbol{\lambda}|\boldsymbol{\rho},\boldsymbol{\Sigma},\mathcal{D} \thicksim\mathcal{N}(\hat{\bs{\lambda}}, \bs{\Omega}^{-1})
$$
where $\bs{\Omega} = \mathbb{I}_{p + q} + \bs{T}'\bs{T}$ and $\hat{\bs{\rho}} = \bs{S}'\bs{T}\bs{\Omega}^{-1}$.
\renewcommand{\theequation}{B-\arabic{equation}} 
\setcounter{equation}{0}  
\section*{Appendix B}
\textbf{Calculation of $CBF_{sec}$}\\
Note that it is immediate from the work above that
$$
\boldsymbol{\rho}|L,\boldsymbol{\lambda},\boldsymbol{\Sigma},\mathcal{D} \thicksim\mathcal{N}_l \left(\hat{\boldsymbol{\rho}}_{L},\boldsymbol{\Xi}_{L}^{-1}\right),
$$
with
\begin{eqnarray*}
\hat{\boldsymbol{\rho}}_{L} &=& \xi^{-1}\tilde{\bs{Y}}'\boldsymbol{V}_{L}\boldsymbol{\Xi}_{L}^{-1}\\
\boldsymbol{\Xi}_{L} &=& \mathbb{I}_{l} + \xi^{-1}\boldsymbol{V}_{L}'\boldsymbol{V}_{L}\\
\boldsymbol{V}_{L} &=& [\bs{X}_{L}\hspace{2mm}\boldsymbol{W}_{L}],
\end{eqnarray*}
where $l$ is the size of model $L$, and $\bs{X}_{L}$ and $\bs{W}_{L}$ denote the columns of the matrices $\bs{X}$ and $\bs{W}$ contained in the model $L$.\\ 
Now, consider $pr(\mc{D}|L,\bs{\lambda},\bs{\Sigma})$.  Note that 
$$
pr(\mc{D}|L,\bs{\lambda},\bs{\Sigma}) = \int_{\Gamma_L} pr(\mc{D}|\bs{\rho},\bs{\lambda},\bs{\Sigma})pr(\bs{\rho}|L)d\bs{\rho}.
$$
Following the calculation in Appendix A, we have
$$
pr(\mathcal{D}|\bs{\rho},\bs{\lambda},\bs{\Sigma})pr(\bs{\rho}|L) \propto (2\pi)^{-l/2}\exp\left(-\frac{1}{2} \left[-2\hat{\bs{\rho}}_{L}\bs{\Xi}_{L}\bs{\rho} + \bs{\rho}'\bs{\Xi}_{L}\bs{\rho}\right]\right).
$$
Substituting this above leads to
$$
pr(\mc{D}|L,\bs{\lambda},\bs{\Sigma})
\varpropto (2\pi)^{-l/2} \int_{\Gamma_L} \exp\left(-\frac{1}{2} \left[-2\hat{\bs{\rho}}_L\bs{\Xi}_L\bs{\rho} + \bs{\rho}'\bs{\Xi}_L\bs{\rho}\right]\right)d\bs{\rho},
$$
which can be expanded to 
\begin{equation}\label{extint}
(2\pi)^{-l/2}\exp\left(\frac{1}{2}\hat{\bs{\rho}}_L'\bs{\Xi}_{L}\hat{\bs{\rho}}_{L}\right)\int_{\Gamma_L} \exp\left(-\frac{1}{2} [\hat{\bs{\rho}}_L'\bs{\Xi}_{L}\hat{\bs{\rho}}_{L}-2\hat{\bs{\rho}}_L'\bs{\Xi}_{L}\bs{\rho} + \bs{\rho}'\bs{\Xi}_L\bs{\rho}]\right)d\bs{\rho}.
\end{equation}
In this form, the integral in \eqref{extint} represents the normalizing constant of a $\mathcal{N}_l(\hat{\bs{\rho}}_{L},\bs{\Xi}_{L}^{-1})$ distribution, i.e.
\begin{eqnarray*} I &=& \int \exp\left(-\frac{1}{2} [\hat{\bs{\rho}}_L'\bs{\Xi}_{L}\hat{\bs{\rho}}_{L}-2\hat{\bs{\rho}}_L'\bs{\Xi}_{L}\bs{\rho} + \bs{\rho}'\bs{\Xi}_{L}\bs{\rho}]\right)d\bs{\rho}\\ 
&=& \frac{|\bs{\Xi}_{L}|^{-1/2}}{(2\pi)^{-l/2}} \int \frac{|\bs{\Xi}_{L}|^{1/2}}{(2\pi)^{l/2}} \exp \left(-\frac{1}{2}[(\bs{\rho}-\hat{\bs{\rho}}_{L})'\bs{\Xi}_L (\bs{\rho}-\hat{\bs{\rho}}_{L})] \right) d\bs{\rho}\\ 
&=& (2\pi)^{l/2}|\bs{\Xi}_{L}|^{-1/2}.  
\end{eqnarray*}
Thus,
$$
pr(\mc{D}|L,\bs{\lambda},\bs{\Sigma}) \propto |\bs{\Xi}_{L}|^{-1/2}\exp\left(\frac{1}{2}\hat{\bs{\rho}}_L'\bs{\Xi}_{L}\hat{\bs{\rho}}_{L}\right).
$$
\noindent \textbf{Calculation of $CBF_{fst}$}\\
Provided that $\rho_1$ is not required to be zero (thus, the endogenous variable is included in model $L$),  we have that
$$
\bs{\lambda}|M,\bs{\rho},\bs{\Sigma},\mathcal{D}\thicksim\mathcal{N} \left(\hat{\bs{\lambda}}_{M},\bs{\Omega}_{M}^{-1}\right),
$$
where \begin{eqnarray*}
\hat{\bs{\lambda}}_{M} &=& \bs{S}'\bs{T}_{M}\bs{\Omega}_{M}^{-1}\\
\bs{\Omega}_{M} &=& \mathbb{I}_{m} + \bs{T}_M'\bs{T}_{M},
\end{eqnarray*}
where $m$ is the size of model $M$ and $\bs{T}_{M}$ is the matrix $\bs{T}$ defined above, but restricted to those variables contained in $M$.\\
\indent When $\rho_1 = 0$, equivalently the endogenous variable is not contained in the model $L$, the approach is altered and essentially becomes a seemingly unrelated regression.  Let $\bs{U}_M = [\bs{Z}_M\mbox{ }\bs{W}_M]$, then we have
$$
\bs{X} = \bs{U}_M\bs{\lambda} + \frac{\sigma_{21}}{\sigma_{11}}\bs{\epsilon}+\bs{\nu}_{2|1},
$$
where, $(\bs{\nu}_{2|1})_i \sim N(0, \omega)$ with $\omega = \sigma_{22} - \sigma_{21}^2/\sigma_{11}$.  Setting $\bs{X}^* = \bs{X} - (\sigma_{21}/\sigma_{11})\bs{\epsilon}$ we have
$$
\bs{X}^* = \bs{U}_M\bs{\lambda} + \bs{\nu}_{2|1},
$$
and by analogy to the steps in Appendix A we see that in this case
\begin{align*}
\hat{\bs{\lambda}}_M &= \omega^{-1}\bs{U}_M'\bs{X}\bs{\Omega}_M^{-1}\\
\bs{\Omega}_M &= \mathbb{I}_{1+p}+\omega^{-1}\bs{U}_M'\bs{U}_M.
\end{align*}
Regardless of how $\hat{\bs{\lambda}}_{M}$ and $\bs{\Omega}_M$ are calculated, the steps in outlining the determination of $CBF_{sec}$ may be followed in this case as well and we see that
\begin{equation*} pr(\mc{D}| M,\bs{\rho},\bs{\Sigma}) \varpropto |\bs{\Omega}_{M}|^{-1/2}\exp\left(\frac{1}{2}\hat{\bs{\lambda}}_M'\bs{\Omega}_{M}\hat{\bs{\lambda}}_{M}\right).
\end{equation*}

\section*{Supplementary Simulation Study}
\indent We conduct a simulation study to evaluate the properties of IVBMA and compare its performance to the Gibbs sampler discussed in Section~\ref{sec:gibbs} that does not incorporate model uncertainty (which we refer to as IV). Our study is similar to that of \citet{lenkoski_et_2011}.\\
\linespread{1}
\begin{table}
{\scriptsize\linespread{1.4}
\begin{center}
\begin{tabular}{l c c c c c c c}
\hline
 & \multicolumn{3}{c}{First Stage}&&\multicolumn{3}{c}{Second Stage}\\
Variable & IV &  IVBMA & True && IV & IVBMA & True\\
\hline
$X$ &   --  &  --  & -- && \textbf{1.498} &  \textbf{1.497} & \textbf{1.5}\\
$W_{1}$ &-0.015 & -0.003  &0 && \textbf{1.986} &  \textbf{1.991} & \textbf{2}\\
$W_{2}$ & \textbf{2.476} &  \textbf{2.480} & \textbf{2.5} && -0.007 &  -0.002 & 0\\
$W_{3}$ & 0.008 &  0.002 & 0 && 0.001 &  0.000& 0\\
$W_{4}$ & -0.008 &  0.001 &0 && \textbf{1.379} &  \textbf{1.384} & \textbf{1.4}\\
$W_{5}$ & -0.007 &  0.001 & 0 && -0.001 &  0.000 & 0\\
$W_{6}$ & 0.004 &  0.000 & 0 && 0.004 & 0.000 & 0\\
$W_{7}$ & 0.000 &  0.000 & 0 && -0.006 &  -0.001 & 0 \\
$W_{8}$ & -0.020 &  -0.005 & 0 && \textbf{2.663} &  \textbf{2.669} & \textbf{2.7}\\
$W_{9}$ & \textbf{1.682} &  \textbf{1.684} & \textbf{1.7} && \textbf{1.226} &  \textbf{1.230} & \textbf{1.25}\\
$W_{10}$ & -0.004 &  -0.003 & 0 && 0.010 &  0.002 &0\\
$W_{11}$ & -0.002 &  0.000 & 0 && 0.006 &  0.003 & 0\\
$W_{12}$ & -0.004 &  -0.002 & 0 && -0.017 &  -0.001 & 0\\
$W_{13}$ & \textbf{0.789} &  \textbf{0.792} & \textbf{0.8} && \textbf{3.265} &  \textbf{3.268} & \textbf{3.3}\\
$W_{14}$ & -0.001 &  0.000 & 0 && -0.010 &  0.001 & 0\\
$W_{15}$ & -0.003 &  0.000 & 0 && -0.004 &  0.000 & 0\\
$Z_{1}$ & 0.011 &  0.001 & 0 &&  --  &  -- & --\\
$Z_{2}$ & -0.001 &  -0.002 & 0 &&  --  &  -- & --\\
$Z_{3}$ & \textbf{4.056} &  \textbf{4.060} & \textbf{4} &&  --  &  -- & -- \\
$Z_{4}$ & -0.004 &  0.001 & 0 &&  --  &  -- & --\\
$Z_{5}$ & 0.002 &  0.000 & 0 &&  --  &  -- & --\\
$Z_{6}$ & 0.002 &  0.001 & 0 &&  --  &  -- & --\\
$Z_{7}$ & \textbf{1.193} &  \textbf{1.195} & \textbf{1.2} &&  --  &  -- & -- \\
$Z_{8}$ & \textbf{2.977} &  \textbf{2.976} & \textbf{3} &&  --  &  -- & --\\
$Z_{9}$ & 0.002 &  0.002 & 0 &&  --  &  -- & -- \\
$Z_{10}$ & \textbf{0.894} &  \textbf{0.896} & \textbf{0.9} &&  --  &  -- & -- \\
\hline
\textbf{MSE} & \textbf{991.92}   & \textbf{541.14} & -- && \textbf{943.79}  & \textbf{541.14} & -- \\
\hline
\end{tabular}
\end{center}
}
\caption{Comparison of parameter estimation under IV and IVBMA across 200 repetitions. Variables shown in bold are those that are included in either the first or the second stage. The values of the total average MSE imply that IVBMA leads to a lower variance in the parameter estimation than IV.}\label{tab:sim_est}
\end{table}
\indent Using the framework in \eqref{equ1}-\eqref{err}, we consider $p=15$ variables in $\boldsymbol{W}$, $q=10$ possible instruments in $\boldsymbol{Z}$ and a univariate endogenous regressor $X$. For simulating data we use $n=120$. These sizes approximately resemble the structure of the data set we will examine in Section \ref{sec:margarine}.  In each synthetic dataset we construct, the values in $\bs{W}$ and $\bs{Z}$ are individually sampled from a $\mc{N}(0,1)$.\\
\indent The variables $\bs{Y}$ and $\bs{X}$ are determined by
\begin{align*}
\bs{Y} &= 1.5\bs{X} + 2\bs{W}_1 + 1.4\bs{W}_4 + 2.7\bs{W}_8 + 1.25\bs{W}_9 + 3.3\bs{W}_{13} + \bs{\epsilon}\\
\bs{X} &= 4.1\bs{Z}_3 + 1.2\bs{Z}_7 + 3\bs{Z}_8 + 0.9\bs{Z}_{10} + 2.5\bs{W}_2 + 1.7\bs{W}_9 + 0.8\bs{W}_{13} + \bs{\eta}.
\end{align*}
Hence, besides $\bs{X}$ five regressors of the vector $\boldsymbol{W}$ have an influence on $\bs{Y}$. Two of these variables also have explanatory power on $\bs{X}$, which is in addition dependent on one further component of $\boldsymbol{W}$, namely $\bs{W}_2$. Finally, only four out of ten candidate variables in $\boldsymbol{Z}$ serve as instruments for $\bs{X}$, while the rest have no explanatory power.\\
\indent Finally, we sample the error terms $\bs{\epsilon}$ and $\bs{\eta}$ from a multivariate normal distribution with a non-diagonal covariance matrix $\boldsymbol{\Sigma}$,
\begin{equation*} 
  \begin{pmatrix}\epsilon_i \\ \eta_i \end{pmatrix}\thicksim \mathcal{N}_{2}\left(0, \begin{pmatrix} 1 &0.4 \\ 0.4 &1\end{pmatrix}\right). 
\end{equation*}
In the following, we use $S=50,000$ as the number of iterations for both methods and discard the first $10,000$ samples as burn-in. The results are averaged over $200$ replications.  Each replication took approximately 45 seconds, on a quad-core 2.8 gHz desktop computer with $16$ GB RAM running Linux.\\
\indent Table \ref{tab:sim_est} displays the results of parameter estimation for the two methods. For each replicate we calculate the posterior expected values $\bar{\bs{\lambda}} = S^{-1}\sum\bs\lambda^{(s)}$ and $\bar{\bs{\rho}} = S^{-1}\sum \bs{\rho}^{(s)}$.  The table then reports the median of these estimates for each variable across the $200$ replicates. Finally, for each replicate we computed the mean squared error (MSE) of the posterior expectations $\bar{\bs{\lambda}}$ and $\bar{\bs{\rho}}$ and report the average of this over all replicates. We can see that for each stage the median of both IVBMA and IV of the posterior expectations are close to the true parameter values.  However, based on the MSE reported in the last row of Table \ref{tab:sim_est} we see that IVBMA leads to considerably lower deviation from the true value than IV estimation. This is because model determination provides a better focus on the variables that have explanatory power on the outcome, which can be seen from the inclusion probabilities shown in Table~\ref{tab:sim_probs}.  This table shows the median and interquartile range of the inclusion probabilities over all $200$ replications.  We see that variables which are included the model are almost always given inclusion probabilities near $1$, while those not in the model typically have very low inclusion probabilities.\\
\vspace{3.5mm}
\linespread{1}
\begin{table}
{\footnotesize\linespread{1.4}
\begin{center}
\begin{tabular}{l c c c c c}
\hline
 & \multicolumn{2}{c}{First Stage}&&\multicolumn{2}{c}{Second Stage}\\
Variable & Median & IQR && Median & IQR\\
\hline
$X$ &  --  &  --  && \textbf{1.000} &  \textbf{(1.000,1.000)}\\
$W_{1}$ & 0.115 &  (0.072,0.493) && \textbf{0.999} &  \textbf{(0.993,1.000)}\\
$W_{2}$ & \textbf{1.000} &  \textbf{(0.991,1.000)} && 0.124 &  (0.083,0.430)\\
$W_{3}$ & 0.107 &  (0.066,0.569) && 0.105 &  (0.069,0.406)\\
$W_{4}$ & 0.110 &  (0.074,0.446) && \textbf{0.999} &  \textbf{(0.997,1.000)}\\
$W_{5}$ & 0.110 &  (0.065,0.371) && 0.098 &  (0.069,0.469)\\
$W_{6}$ & 0.101 &  (0.068,0.599) && 0.102 &  (0.066,0.518)\\
$W_{7}$ & 0.101 &  (0.064,0.435) && 0.101 &  (0.068,0.403)\\
$W_{8}$ & 0.115 &  (0.074,0.560) && \textbf{1.000} &  \textbf{(0.995,1.000)}\\
$W_{9}$ & \textbf{0.999} &  \textbf{(0.991,1.000)} && \textbf{0.999} &  \textbf{(0.993,1.000)}\\
$W_{10}$ & 0.106 &  (0.067,0.521) && 0.107 &  (0.069,0.507)\\
$W_{11}$ & 0.102 &  (0.067,0.516) && 0.111 &  (0.070,0.447)\\
$W_{12}$ & 0.104 &  (0.067,0.464) && 0.101 &  (0.070,0.458)\\
$W_{13}$ & \textbf{0.999} &  \textbf{(0.995,1.000)} && \textbf{1.000} &  \textbf{(0.996,1.000)}\\
$W_{14}$ & 0.101 &  (0.070,0.475) && 0.099 &  (0.070,0.458)\\
$W_{15}$ & 0.096 &  (0.068,0.313) && 0.101 &  (0.073,0.308)\\
$Z_{1}$ & 0.103 &  (0.065,0.403) &&  --  &  -- \\
$Z_{2}$ & 0.105 &  (0.071,0.507) &&  --  &  -- \\
$Z_{3}$ & \textbf{0.999} &  \textbf{(0.987,1.000)} &&  --  &  -- \\
$Z_{4}$ & 0.109 &  (0.071,0.569) &&  --  &  -- \\
$Z_{5}$ & 0.097 &  (0.065,0.483) &&  --  &  -- \\
$Z_{6}$ & 0.100 &  (0.068,0.722) &&  --  &  -- \\
$Z_{7}$ & \textbf{0.999} &  \textbf{(0.984,1.000)} &&  --  &  -- \\
$Z_{8}$ & \textbf{0.999} &  \textbf{(0.988,1.000)} &&  --  &  -- \\
$Z_{9}$ & 0.104 &  (0.070,0.532) &&  --  &  -- \\
$Z_{10}$ & \textbf{0.999} &  \textbf{(0.985,1.000)} &&  --  &  -- \\
\hline
\end{tabular}
\end{center}
}
\caption{Median and IQR of variable inclusion probabilities across 200 repetitions.  Variables shown in bold are those that are included in either the first or the second stage.  This table shows that inclusion probabilities closely match the true structure of the system.}\label{tab:sim_probs}
\end{table}

\end{document}